\documentstyle[prl,twocolumn,aps]{revtex}
\input{epsf}
\begin{document}
\draft
\twocolumn[\hsize\textwidth\columnwidth\hsize\csname @twocolumnfalse\endcsname
\title{Quantum Chaos $\&$ Quantum Computers}

\author{D. L. Shepelyansky$^{^{*}}$}

\address {Laboratoire de Physique Quantique, UMR 5626 du CNRS, 
Universit\'e Paul Sabatier, F-31062 Toulouse Cedex 4, France}

\date{June 16, 2000}

\maketitle

\begin{abstract}
The standard generic quantum computer model is studied
analytically and numerically and the border
for emergence of quantum chaos, induced by imperfections
and residual inter-qubit couplings, is determined.
This phenomenon appears in an isolated quantum computer
without any external decoherence.
The onset of quantum chaos leads to
quantum computer hardware melting,
strong quantum entropy growth and
destruction of computer operability.
The time scales for development of quantum chaos and ergodicity
are determined.
In spite the fact that this phenomenon
is rather dangerous for quantum computing
it is shown that the quantum chaos border for
inter-qubit coupling 
is exponentially larger than the energy level spacing
between quantum computer eigenstates and drops only
linearly with the number of qubits $n$. 
As a result the ideal multi-qubit structure 
of the computer remains
rather robust against imperfections.
This opens a broad parameter region for a possible 
realization of quantum computer.
The obtained results are related to the recent
studies of quantum chaos in such
many-body systems as nuclei, complex atoms and molecules,
finite Fermi systems and quantum spin glass shards
which are also reviewed in the paper.
\end{abstract}
\large {Lecture at Nobel symposium on ``Quantum chaos'', 
June 2000, Sweden}
\vskip1pc]

\narrowtext
\section{Introduction}
On the border between two Millennia it is natural to ask
a question, {\it what will be the origin of future  human power?}
Even thirty or twenty years ago the standard
answer would be: nuclear. But now in a view of
amazing computer development all over the world
it becomes clear that the future power will be related to ability
to count as fast as possible. In I Millennium this ability
was basically comparable with finger counting,
while at the end of II Millennium  it made
enormous jump with computer creation 
which led to a qualitative change in 
human society. During last two decades
the power of modern computers
demonstrated a constant impressive growth due
to technological progress and creation of chips
of smaller and smaller size. In a near future
this size  should reach a scale at which the quantum
nature of physical laws will become dominant.
As a result, we unavoidably come to the creation problem
of quantum computer. Such a computer should be essentially 
based on quantum mechanics and
operate with unitary transformations and quantum logic.
The unitary nature of transformations allows to exclude
energy dissipation that should play an important role
on small scales. At the same time, as stressed 
by Feynman \cite{feynman}, the classical computer 
has enormous difficulty  in simulation of many-body
quantum systems due to exponential growth of the Hilbert space
with the number of particles and hence, of the computational
efforts. Due to that it is possible to expect that
a computer composed of quantum elements will be
much more efficient for 
solution of quantum, and may be other, problems.
At present quantum computer is viewed as a system of coupled
$n$ qubits being two-level quantum systems or one-half spins
(see recent review \cite{steane2} and references there in).
The computer operation is based on controlled series
of two-qubit coupling switch on and off  which together with 
one-qubit rotations allow to realize any unitary operation
in the Hilbert space of size $N_H=2^n$ \cite{deutsch,divi2}.
In this respect the inter-qubit coupling becomes  unavoidable
property of quantum computer.
 
Recently a great increase of interest to quantum computing
has been generated by the work of Shor \cite{shor1}
who constructed a quantum
algorithm which performs large number factorization 
into primes exponentially
faster than any known classical algorithm.
Also Grover showed \cite{grover} that a search of an item
in a long list is done much faster by quantum computer.
The enormous gain in the computation rate
is reached due to high parallelism 
of multi-qubit quantum evolution and 
quantum interference.
Together with a recent theoretical development of
quantum error-correcting codes \cite{shor2,steane1}
these exciting results stimulated various experimental proposals
for realization of quantum computer.
The variety of physical systems proposed is really amazing
and includes: ion traps \cite{zoller}, nuclear magnetic resonance 
systems \cite{nmr}, nuclear spins with interaction controlled 
electronically \cite{vagner,kane} or by laser pulses \cite{bowden}, 
quantum dots \cite{loss}, Cooper pair boxes \cite{cooper}, 
optical lattices \cite{lattice} and electrons floating on liquid 
helium \cite{helium}. At present two-qubit gates were
experimentally realized with cold atoms \cite{monroe}, and 
the Grover algorithm has been performed for 
three qubits made from nuclear spins in a molecule \cite{3q}.

Thus there are two main lines in the present day
quantum computer research:
construction and development of efficient quantum algorithms
and search for a optimal physical system with a future
experimental realization of few coupled
qubits. The first line has a strong mathematical shade:
indeed, it assumes that all qubits are perfectly
identical and the couplings between them can be 
operated also in a perfect way. The second line, in a large
respect, is a part of experimental physics with few  qubits.
As a result, there is a broad open field for
physical studies of a realistic quantum computer
with many qubits. 
Indeed, in reality the qubits are never perfect,
the level spacing fluctuates from one qubit to another
due to different environment, there is also a
residual interaction between qubits which
cannot be eliminated completely, the two-qubit gate operations
are also not perfect. In practice, at least around $n = 1000$
of such qubits are required to make a quantum computer
more efficient than the existing our days computers \cite{steane2}.
In addition to the above internal
imperfections there is also decoherence due to coupling to
external world  which produces noise and
dissipation. The effects of decoherence and two-qubit
gates pulse  broadening were numerically tested
on Shor's algorithm \cite{paz,zurek} and were shown to
play an important role. At the same time the estimates
show that it is possible to have physical qubits
with very long relaxation time,
during which many gate operations can be realized \cite{divi1}.
One of the  most promising system looks to be 
nuclear spins in two-dimensional semiconductor structures
\cite{vagner,kane,vagner2}.

However, the absence of external decoherence does not yet
mean that the computer will operate properly.
Indeed, internal imperfections with inter-qubit residual 
couplings $J$ can strongly modify the ideal quantum register
represented by noninteracting many-body (multi-qubit) states
of ideal qubits. A simple estimate \cite{gs1} shows that the
residual interaction $J$ will be unavoidably much larger
than the energy level spacing $\Delta_n$ between adjacent 
eigenstates of a realistic/generic quantum computer .
Let us assume  that $J$ is relatively weak comparing to
one-qubit level spacing $\Delta_0$. Then all $N_H$ eigenenergies
will be located in the energy band of size 
$\Delta E \sim n \Delta_0$ and the average multi-qubit 
level spacing is
$\Delta_n \approx \Delta E/N_H \sim \Delta_0 2^{-n} \ll \Delta_0$.
For the experimental proposals \cite{vagner,kane}
we have $\Delta_0 \sim 1$ K so that for $n=1000$,
when Shor's algorithm becomes useful, the multi-qubit spacing
is incredibly small 
$\Delta_n \sim  10^3 \times 2^{-10^3} \Delta_0 \sim 10^{-298}$ K.
This value will definitely
be much smaller than any physical residual interaction $J$.  For
the proposal \cite{kane}  with a distance
between nuclear spins of $r=200 $ {\AA} and an effective 
Bohr radius of $a_B=30$ {\AA} ( Eq.(2) of \cite{kane}), 
the coupling between qubits
(spin-spin interaction)
is $J \sim \Delta_0 \sim 1$ K. By changing the electrostatic gate potential,
an effective barrier between nuclei can be modified
that can be viewed as a change of
effective electron mass possibly up to a factor of two.
Since $J  \propto (r/a_B)^{5/2} 
\exp (-2r/a_B)/a_B$, and $a_B$ is inversely proportional to the effective mass,
this gives a minimal residual spin-spin 
interaction of $J \sim 10^{-5}$ K $\gg \Delta_n$.  
On the first glance this would lead to a natural/naive
conclusion that at such residual interaction the ideal 
quantum computer eigenstates are strongly mixed and 
completely modified resulting in destruction of
quantum computer hardware. In spite of this expectation
it has been shown recently that the ideal qubit structure 
is much more robust and in reality the quantum 
hardware melting and quantum chaos induced by
inter-qubit interaction takes place at  $J > J_c \sim \Delta_0/n $ being
exponentially larger than $\Delta_n$ \cite{gs1}. This result
for quantum chaos border in quantum computing
is recently confirmed by more extended studies \cite{gs2}
and opens a broad regime of parameters for which
realization of a quantum computer is possible.
For example, at $\Delta_0 \sim 1$ K and $n=1000$
the critical coupling $J_c \sim 1$ mK is compatible with the
experimental proposal \cite{kane}.

The above result is closely related to the long term
research  of quantum many-body
systems, started by Wigner \cite{wigner} interested in 
``the properties of the wave functions of quantum 
mechanical systems which are assumed to be so complicated 
that statistical consideration can be applied to them''. 
As a result,  the random matrix
theory (RMT) has been developed to explain the generic
properties of complex eigenstates and energy spectra
of many-body interacting systems such as
heavy nuclei, many electron atoms and molecules 
\cite{mehta}. The success of RMT was so 
impressive \cite{guhr} that the conditions
of its applicability to many-body systems 
were not really realized until recently.
Indeed, for example, in nuclei the density of states 
grows exponentially with excitation energy and at a
first glance the RMT is valid
as soon as the interaction matrix elements $U$ become
comparable to the level spacing between many-body
states $\Delta_n$. However, in nature we have only
two-body interaction and therefore, while the size
of the Hilbert space grows exponentially with
the number of particles $n$, the number of nonzero
interaction induced matrix elements grows not 
faster than $n^2$. To study the spectral statistics
in this situation a two-body random interaction model
(TBRIM) was introduced 
and it was shown \cite{tbrim1,tbrim2} 
that in the limit of strong
interaction the RMT remains valid even if
the full Hamiltonian matrix is exponentially 
sparsed. However, much more time was needed to
understand the case of relatively weak interaction
and to find the critical $U_c$ above which
quantum chaos and RMT set in. Contrary to
naive expectation $U \sim \Delta_n$ 
assumed by many authors until recently 
(see e.g. \cite{flambaum1,zelevin,berkovits}) 
it now became clear that
the quantum chaos border is exponentially larger than
$\Delta_n$ since only transitions between
directly coupled states, and not the bulk density of states, 
are important for level mixing. 
As a result the interaction should be larger
than the energy level spacing $\Delta_c$
between directly coupled states $U > U_c  \sim \Delta_c \gg \Delta_n$
to change the level spacing statistics $P(s)$
from the Poisson distribution to the Wigner-Dyson one
and to generate quantum ergodicity and chaos for
eigenstates. As for my knowledge, for the first
time this condition for quantum chaos in many-body
systems had been formulated by  {\AA}berg 
who also by numerical simulations for a nucleus model
had found the change in the level statistics 
across this border \cite{aberg}. 
Due to that  I will call this condition {\it {\AA}berg criterion}.

In spite of other {\AA}berg's papers (e.g. \cite{aberg1})
his result was not broadly known for the community
probably because his studies were mainly 
addressed to nuclear physicists and the numerical
results were not so confirmatory at that time.
Also in nuclei the interaction is usually quite strong
and the quantum chaos sets in rather quickly.
In fact it was shown by numerical simulations
of other models that in an isolated
many-body Fermi system a sufficiently strong interaction
can induce dynamical thermalization with 
the Fermi-Dirac distribution \cite{flambaum1,zelevin}.
Independently , the line of research related to
the problem of two interacting
particles in a random potential \cite{dls94}
showed that a two-body interaction can lead to 
a number of unexpected results and that disorder/chaos
can strongly enhance the interaction. The last
property in fact had been known from the spectacular
Sushkov-Flambaum enhancement of weak
interaction in nuclei \cite{sushkov}. The analytical
studies of few particles with two-body random interaction $U$ 
\cite{dlssush97} showed that the mixing of levels, quantum chaos 
and RMT statistics appear only if $U > \Delta_2$,
where $\Delta_2 \gg \Delta_3 \gg \Delta_4 ...$
are  the  level spacing between 2-,3-,4-particle levels
respectively. This result was generalized for many
particles and the quantum chaos border $U_c \sim \Delta_c$
was proposed and confirmed by extensive numerical
studies of the TBRIM model \cite{jacquod97},
independently of {\AA}berg's papers.
As it will be seen from the present paper the knowledge
obtained for quantum many-body systems 
can be successfully used for such a new direction
of research as quantum computing. 

In this paper I review the recent developments of
quantum chaos theory in many-body systems
obtained in Toulouse  
and show their links and importance for a
quantum computer which also represents a many-body
system with exponentially large Hilbert space.
The paper is organized as follows. In the next Section
the analytical and numerical results are presented
for emergence of quantum chaos in many-body systems.
In Section III the standard generic quantum computer
(SGQC) model is described and the results
of Section II are generalized for 
quantum chaos border for quantum computing.
In Section IV the time evolution in the regime of
quantum chaos is studied and different time scales
imposed  by chaos and decoherence 
on quantum computing are discussed.
The paper ends by the concluding remarks in the
last Section. 

\section{{\AA}berg criterion for emergence of quantum chaos
in many-body systems}

Following \cite{dlssush97} let us consider first a case
of three particles located on $m $ one-particle orbitals
with energy level spacing $\Delta \sim V/m$.
For simplicity we assume the particles 
to be  distinguishable that however is not of 
principle importance for $m \gg 1$. 
The level spacing between 2,3-particle states
is $\Delta_2 \sim V/m^2, \; \Delta_3 \sim V/m^3$ respectively 
with $\Delta \gg \Delta_2 \gg \Delta_3$.
The two-body matrix elements
written in the noninteracting eigenbasis 
are supposed to be random with a typical
value $U_{12} \sim U$ and $U_{23} \sim U$
for interaction between 1st/2d and 2d/3d
particles respectively. At the same time
the interaction matrix element $U_{13}$ between
1st/3d is taken to be zero for simplicity
(the final result remains the same 
for $U_{13} \sim U$). For two interacting
particles, e.g. 1st/2d, the levels
become mixed by interaction and RMT spectral statistics sets in
when the interaction becomes larger than the
level spacing  between two-particles states:
$U_{12} > \Delta_2$. On the contrary for 
$U_{12} \ll \Delta_2$ the perturbation theory 
is valid and the eigenstates with interaction
are determined by one noninteracting
eigenstate. In the case of three particles
the situation is more complicated since 
the three-particle states are not directly 
coupled by two-body interaction. 

The matrix element $U_3$ between 3-particle
levels can be found by the second order perturbation
theory. In this way the matrix element
between initial state $|123>$ and final state $|1'2'3'>$
is given by diagram presented in Fig.1 
with intermediate state $|1'{\bar 2} 3>$.
\begin{equation}
\label{3part}
U_{3} = {\sum_{\bar 2}} 
{{{<12| U_{12} |1' {\bar 2}> < {\bar 2} 3| U_{23}| 2' 3'>} \over
{(E_1+E_2+E_3 - E_{1'}-E_{\bar 2}-E_3)}}}
\sim  {{U^2} \over  {\Delta}}
\end{equation}
\noindent
It is important that the summation is carried out only over single particle
states $\bar 2$ and the sum is mainly determined by a term
with a minimal detuning in denominator being of order $\Delta$.
As a result the level mixing sets in
for $U_3 > \Delta_3$ that  gives the quantum chaos border \cite{dlssush97}:
\begin{equation}
\label{3pborder}
U \sim U_c \sim \Delta_2 \gg \Delta_3
\end{equation}
This means that the 3-particle levels are mixed only when
the interaction mixes two-particle levels that is the consequence
of the  two-body nature of interaction.
In a similar way for few particles $n \sim 3$
the border $U_c \sim \Delta_2$. This conclusion was
confirmed in \cite{weinmann}.
\begin{figure}
\epsfxsize=2.8in
\epsfysize=2.2in
\epsffile{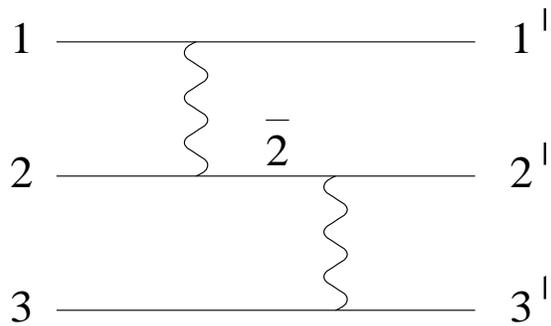}
\vglue -0.2cm
\caption{Diagram for the effective 3-particle 
matrix element $U_3$ in 
(\ref{3part}), after [38].
} 
\label{fig1}
\end{figure}

Let us now consider a more general case of TBRIM 
\cite{tbrim1,tbrim2,flambaum1} with $n$ 
Fermi particles distributed over $m$ energy orbitals
$\epsilon_{m'}$, $m'=1,2,...,m$. These energies are randomly
and homogeneously distributed in the interval
$[0, m\Delta]$ with spacing $\Delta$. The total number of many-body 
states is $N=m!/(n!(m-n)!)$ and they are coupled by 
random two-body matrix elements which value is
uniformly distributed in the interval $[-U,U]$.
Due to the two-body nature of interaction
the number of multi-particle states coupled by interaction,
or the number of direct transitions,  is
$K=1+n(m-n)+n(n-1)(m-n)(m-n-1)/4$ \cite{flambaum1}.
All these transitions occur inside a two-body energy interval 
$B=(2m-4) \Delta$ around the energy of 
initial multi-particle state. 
For large $m$ and $n$ the number of transitions
$K$ is much smaller than the size of the matrix $N$ but  is much
larger than the number of different two-body matrix elements 
$N_2 \approx m^2/2$. The Fermi energy of the systems is 
$\epsilon_F \approx n \Delta$ and the level spacing
in the middle of the total energy band is exponentially small
$\Delta_n \approx (m-n)n\Delta/N$.
\begin{figure}
\epsfxsize=3.4in
\epsfysize=2.6in
\epsffile{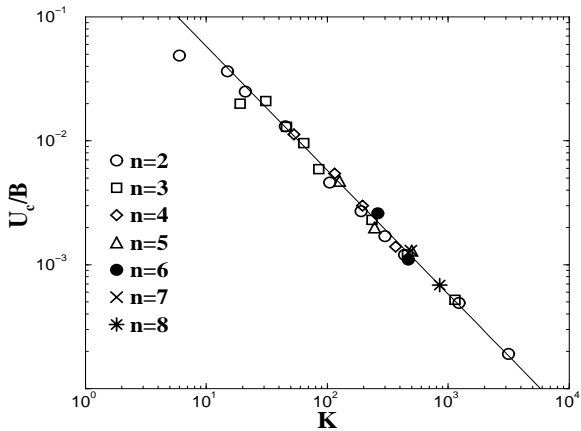}
\vglue 0.2cm
\caption{Dependence of the rescaled critical interaction strength $U_c/B$,
above which $P(s)$ becomes close to the Wigner-Dyson statistics,
on the number of directly coupled states $K$ for $4 \leq m \leq 80$
and $1/40 \leq n/m \leq 1/2$. The line shows the theory (\ref{uc}) with
$C=0.58$, after [40].
} 
\label{fig2}
\end{figure}

At the middle of spectrum the density of {\it directly coupled}
states is $\rho_c=K/B$. According to the usual perturbation theory,
these states, if they would be alone, 
are  mixed if the coupling $U$ is larger than their level
spacing $\Delta_c=1/\rho_c$.
This is the {\AA}berg criterion \cite{aberg,aberg1},
which was also independently proposed in
\cite{jacquod97}. This relation is also in agreement with
the arguments given above for few particles.
As a result, the onset of quantum chaos 
is expected for $U>U_c$: 
\begin{equation}
\label{uc}
U_c = C \Delta_c=C \frac{B}{K} \approx \frac{2 C}{\rho_2 n^2}
\end{equation}
where $C$ is a numerical constant to be found
and $\rho_2 \approx N_2/B \approx m/4\Delta$ is 
the two-particle density at $m \gg n \gg 1$.
It is important to stress that the critical
coupling $U_c$ is exponentially larger than the level
spacing $\Delta_n$ between multi-particle states.
\vskip -0.5cm
\begin{figure}
\epsfxsize=3.4in
\epsfysize=2.6in
\epsffile{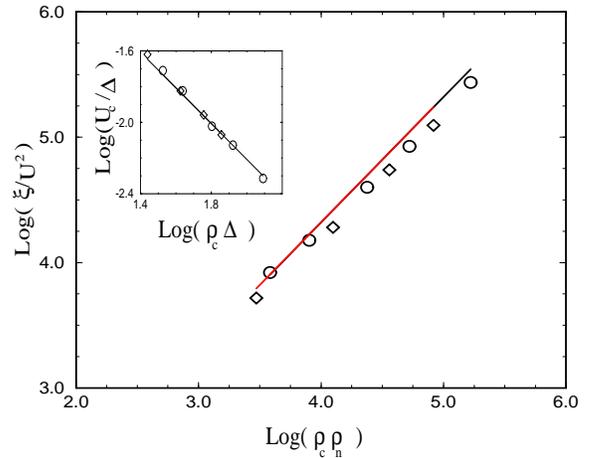}
\vglue 0.2cm
\caption{Dependence of the rescaled IPR $\xi /U^2$  on 
$\rho_c \rho_n$: layer model data for $n=3$, $\Delta=1$
and $40\leq m \leq 130$ (o);
$n=4$ and $30\leq m \leq 60$ (diamonds).  The straight line gives
theory (\ref{xi}).   Insert shows $ U_c/\Delta$ vs.
$\rho_c \Delta$ in log-log scale for the same parameters; 
the straight line is the fit $U_c = 0.62/\rho_c$.
After [43].
} 
\label{fig3}
\end{figure}

The most direct way to detect the emergence of quantum chaos
is by the change in the probability
distribution $P(s)$ of nearest level spacings $s$,
where $s$ is measured in units of average spacing.
Indeed, for integrable systems the levels are noncorrelated
and characterized by the Poisson distribution $P_P(s)=\exp(-s)$,
while in the quantum chaos regime the statistics is close to the 
Wigner surmise $P_W(s)=(\pi s/2)\exp(-\pi s^2/4)$ \cite{bohigas}.
To identify a transition from one limiting case to another
it is convenient to introduce the parameter 
$\eta = \int_0^{s_0}
(P(s)-P_{W}(s)) ds / \int_0^{s_0} (P_{P}(s)-P_{W}(s)) ds$,
where $s_0=0.4729...$ is the intersection point
of $P_P(s)$ and $P_W(s)$. In this way $\eta$ varies from 1
($P(s)=P_P(s)$) to 0 ($P(s)=P_{W}(s)\;$) and the critical 
value of $U_c$ can be determined by the condition
$\eta(U_c)=\eta_c=0.3$ \cite{jacquod97}. In fact the choice of $\eta_c$
affects only the numerical constant $C$ in (\ref{uc}).
The chosen $\eta_c=0.3$ is close  to the value
$\eta_c \approx 0.2$ for the critical
statistics at the Anderson transition on a
3-dimensional disordered lattice \cite{shklovskii}.
The results of extensive numerical studies, performed in
\cite{jacquod97}, are presented in Fig.2 and 
confirm the {\AA}berg 
criterion (\ref{uc}) in a large parameter range
with $C=0.58$ \cite{anote}. Of course, the data \cite{jacquod97}
are obtained for a much larger parameter range than 
in \cite{aberg,aberg1}. 
Nevertheless, the  above $C$ value is in a good
agreement with a numerical factor found in \cite{aberg1}
(see Eq. 22 there), which gives $C \approx 0.7$,
if to take into account that $C$ is defined via
the same average values of square matrix element.
A similar value is also found in more advanced 
studies for $n=3,4$ in the layer model approximation
with the states selected in the energy interval $\Delta$
(see insert in Fig. 3)\cite{georgeot} . 
Also the above studies \cite{aberg,aberg1,jacquod97,georgeot}
show that contrary to the sharp Anderson transition 
\cite{shklovskii} a smooth crossover from one statistics to 
another takes place at $U \approx U_c$
(see, however, \cite{song} and references therein).

The {\AA}berg criterion can be applied not only to 
excited states but also to low energy excitations 
near the Fermi level. Suppose we have Fermi gas
with a temperature $T \ll \epsilon_F$. Then,
according to the Fermi-Dirac distribution,
the number of effectively interacting particles
is $\delta n \sim Tn/\epsilon_F$ with
the density of these two-particle states 
$\rho_2 \sim T/\Delta^2$
and $\rho_c \sim \rho_2 {\delta n}^2 \sim \Delta (T/\Delta)^3$. 
The total excitation energy is 
$\delta E \sim T \delta n \sim T^2/\Delta$.
As a result the interaction induced/dynamical
thermalization and the quantum chaos set in \cite{jacquod97} only for
\begin{equation}
\label{Tc}
\delta E > \delta E_{ch} \approx \Delta (\Delta/U)^{2/3}; \;\;\;
T > T_{ch} \approx  \Delta (\Delta/U)^{1/3}
\end{equation}
These relations follow also from the {\AA}berg 
papers \cite{aberg,aberg1} even if they were not 
written directly there. 
The numerical constant in (\ref{Tc}) corresponds to 
$\eta_c=0.3$ \cite{jacquod97}. Below the border (\ref{Tc})
the eigenstates are not ergodic and the interaction is
too weak to thermalize the fermions even if the multi-particle
level spacing is exponentially small 
($\Delta_n \propto \exp(-2.5 (\delta E/ \Delta)^{1/2}$
\cite{bohr}). After \cite{aberg,aberg1,jacquod97}
the dependence (\ref{Tc}) was also obtained
in \cite{mirlin97}.

In the quantum chaos regime $U>U_c$ the local density of states
is described by the Breit-Wigner distribution
with the energy width $\Gamma$ 
given by the Fermi golden rule $\Gamma = 2 \pi U^2 \rho_c/3$
\cite{georgeot}. The value of $\Gamma$ determines 
the spreading width of eigenstates mixed by interaction.
The number of noninteracting eigenstates
contributing to a given eigenstate can be measured
through the inverse participation ratio (IPR)
$\xi = 1/{\sum_i |a_i|^4}$, where $a_i$ are probability 
amplitudes in the noninteracting eigenbasis. 
The mixing of all levels in the interval $\Gamma$
gives \cite{georgeot}
\begin{equation}
\label{xi}
\xi \approx \Gamma \rho_n \approx 2 U^2 \rho_c \rho_n.
\end{equation} 
where the numerical factor is taken in analogy with
the known result for band random matrices.
This analytical relation is in a good agreement 
with the numerical data shown in Fig. 3.
It is important to note that at $U=U_c$ exponentially
many states are mixed by interaction. The width $\Gamma$ 
has also another important physical meaning: it determines
the chaotic time scale $\tau_{\chi} \approx 1/\Gamma$
after which an initial noninteracting eigenstate
disintegrates over exponentially many ($\xi$)
eigenstates of interacting system (here and below $\hbar=1$). 

\section{Standard generic quantum computer model}

In \cite{gs1} the standard generic quantum computer (SGQC) model was introduced
to describe a system of $n$ qubits containing imperfections which generate
a residual inter-qubit coupling and fluctuations in the
energy spacings between the two states of one qubit.
The Hamiltonian of this model reads:
\begin{equation}
\label{hamil}
H = \sum_{i} \Gamma_i \sigma_{i}^z + \sum_{i<j} J_{ij} 
\sigma_{i}^x \sigma_{j}^x,
\end{equation}
where the $\sigma_{i}$ are the Pauli matrices for the qubit $i$ and the second
sum runs
over nearest-neighbor qubit pairs on a two-dimensional lattice
with periodic boundary conditions applied.
The energy spacing between the two states of a qubit is represented 
by $\Gamma_i$ randomly and uniformly distributed in the interval 
$[\Delta_0 -\delta /2, \Delta_0 + \delta /2 ]$.  The detuning
parameter $\delta$ gives the width of the distribution 
near the average value $\Delta_0$ 
and may vary from $0$ to $\Delta_0$.  Fluctuations
in the values of $\Gamma_i$ appear generally as a result of imperfections,
e.g. local magnetic field and density fluctuations in 
the experimental proposals \cite{vagner,kane}.
The couplings $J_{ij}$ represent the residual static interaction 
between qubits which
is always present for reasons explained in the introduction.
They can originate from spin-exciton exchange \cite{vagner,kane},
Coulomb interaction \cite{zoller}, 
dipole-dipole interaction \cite{helium}, etc...
To catch the general features of the different proposals,
$J_{ij}$ are chosen randomly and uniformly distributed 
in the interval $[-J,J]$. This SGQC model describes
the quantum computer hardware, while the gate operation in time should
include additional time-dependent terms in the Hamiltonian (\ref{hamil})
and will be studied separately.  At $J=0$ the noninteracting
eigenstates of the
SGQC model  can be presented as $|\psi_i>
=|\alpha_1,...,\alpha_n>$ where $\alpha_k=0,1$ marks the polarization
of each individual qubit.  These are the ideal eigenstates of a quantum
computer called  quantum register states.  For $J \neq 0$,
these states are no longer eigenstates of the
Hamiltonian, and the new eigenstates are now 
linear combinations of different quantum register states.  
The term multi-qubit states is used to denote the eigenstates 
of the SGQC model with interaction but also for the case $J=0$.
It is important to stress that the quantum computer
operates in the middle of energy spectrum where the density
of states is exponentially large and it is natural
that the quantum chaos initially sets in
this bulk part of the spectrum. In this respect low energy excitations
are not important contrary to fermionic systems
discussed above.

It is interesting to note that when one site in (\ref{hamil})
is coupled with all other sites and $\delta=2 \Delta_0$ then
the system becomes equivalent to the quantum version of the classical
Sherrington-Kirpatrick spin glass model studied in \cite{glass}.
For such a quantum spin glass shard
in the middle of the spectrum the coupling matrix
element is $U \sim J$, the spacing between directly coupled states
is $\Delta_c \sim \Delta_0/n^2$, since each state is coupled 
to $ n(n-1)/2$ states in the energy band of the order of
$\Delta_0$. Therefore, according to (\ref{uc})
the quantum chaos and ergodicity emerge for
$J > \Delta_0/n^2 \gg \Delta_n$ as it was shown
analytically and numerically in \cite{glass}.   

A similar analysis can be done for the SGQC model (\ref{hamil}).
Indeed, for $\delta \ll \Delta_0$ and $J < \delta$
the total spectrum is composed
of $n$ bands with inter-band spacing $2\Delta_0$ and band width 
$\sqrt{n} \delta$. Within one band one quantum register state
is coupled to about $n$ states in an energy window
of $2\delta$ so that $\Delta_c \sim \delta/n$ and the quantum 
chaos border is given by \cite{gs1}
\begin{equation}
\label{qcborder}
J_c = C_q \delta/n
\end{equation}
where $C_q$ is a numerical factor. All above arguments
remain valid up to $\delta=\Delta_0$ when the bands
become overlapped. It is important that $J_c \gg \Delta_n$
and that $J_c$ decreases with $\delta$. The last property is
natural since at small $\delta$ the states in one band 
are quasi-degenerate and it is easier to mix levels. 

\begin{figure}
\epsfxsize=3.4in
\epsfysize=2.6in
\epsffile{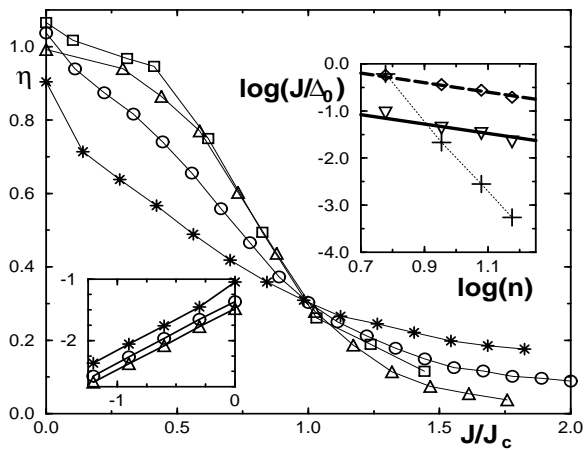}
\vglue -0.2cm
\caption{Dependence of $\eta$ on the rescaled coupling strength $J/J_{c}$
for the states in the middle of the energy band  for
$n=6 (*),9 $(o)$,12 $(triangles)$,15$(squares); 
$\delta=\Delta_0$.
The upper insert shows $\log(J_c/\Delta_0)$ (diamonds) 
and $\log(J_{cs}/\Delta_0)$ (triangles) 
versus $\log(n)$; the variation of the scaled multi-qubit
spacing $\Delta_n/\Delta_0$ with $\log(n)$ is shown for comparison (+).
Dashed line gives the theoretical formula 
(\ref{qcborder}) with $C_q=3.16$; the solid line is $J_{cs}=0.41 \Delta_0/n$.
The lower insert shows $\log(J_{cs}/\Delta_0)$ versus $\log(\delta/\Delta_0)$
for $n=6 (*),9$ (o), 12 (triangles); straight lines have
slope $1$. After [24].
}
\label{fig4}
\end{figure}

The direct numerical simulations for the quantum chaos
border in quantum computer are done in  \cite{gs1} for 
the SGQC model. The change in the level spacing statistics 
$P(s)$ in the band center 
with the growth of residual interaction $J$
can be quantitatively characterized by the parameter
$\eta$. To suppress fluctuations $P(s)$ is obtained
by averaging over $5 \leq N_D \leq 4 \times 10^4$
random realizations of $\Gamma_i, J_{ij}$ so that
the total spacing statistics was 
$10^4 < N_S \leq 1.6 \times 10^5$. Also $P(s)$
is determined inside one of the symmetry classes
of (\ref{hamil}) with odd or even number of qubits up. 
The dependence of $\eta$ on $J$ is presented in Fig. 4.
The variation of critical coupling with $n$ and $\delta$
can be determined from the condition
$\eta(J_c)=0.3$. The data obtained are in a good
agreement with (\ref{qcborder}) with $C_q=3.16$
and clearly show that $J_c \gg \Delta_n$ (see inserts
in Fig. 4). According to Fig. 4 the transition is sharp
in the limit of large $n$ in contrast to the smooth crossover
in the TBRIM. This difference is due to local
interaction between particles in (\ref{hamil})
contrary to the long range interaction in the TBRIM.
  
In the limit $\delta \ll \Delta_0$ and $J \ll \Delta_0$
the coupling between different energy bands is 
negligibly small. In this case to a good approximation 
the SGQC Hamiltonian (\ref{hamil}) can be reduced to 
the renormalized Hamiltonian  
$H_{P}=\Sigma_{k=1}^{n+1} \hat{P_k} H \hat{P_k}$ where $\hat{P_k}$
is the projector on the $k^{th}$ band, so that qubits are coupled only
inside one band. For an even $n$ this band is centered exactly at $E=0$,
while for odd $n$ there are two bands centered at $E=\pm \Delta_0$,
and we will use the one at $E=-\Delta_0$.  Such a band corresponds
to the highest density of states, and in a sense represents 
the quantum computer core. The dependence of
critical coupling, determined via $\eta(J_c)=0.3$,
on the number of qubits $n$ is shown in Fig. 5 
being in  good agreement with the theoretical quantum chaos
border (\ref{qcborder}).  It is important to note that
at $\delta=0$ the parameter $\eta=0$ and
the eigenstates are chaotic \cite{gs2} in agreement with
(\ref{qcborder}).

\begin{figure}
\epsfxsize=3.3in
\epsfysize=2.6in
\epsffile{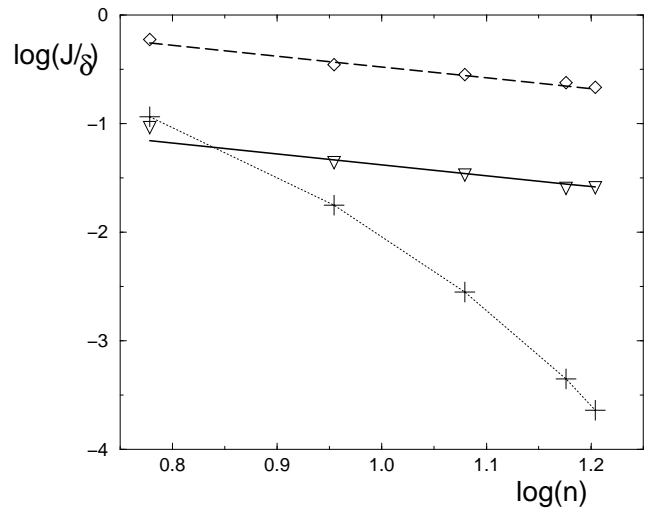}
\vglue 0.2cm
\caption{Dependence of $\log(J_c/\delta)$ (diamonds) 
and $\log(J_{cs}/\delta)$ (triangles) 
versus $\log(n)$; the variation of the scaled multi-qubit
spacing ($\log(\Delta_n/\delta))$ with $\log(n)$ is shown for comparison (+).
Dashed line gives the theoretical formula 
(\ref{qcborder}) with $C_q=3.3$; the solid line is $J_{cs}=0.41 \delta/n$;
the dotted curve is drawn to guide the eye for (+).
After [25].
}
\label{fig5}
\end{figure}

The drastic change in the level spacing statistics
is in fact related to a qualitative change in the
quantum computed eigenstate structure. For $J \ll J_c$
the eigenstates are very close to noninteracting
multi-qubit states $|\psi_i>$,  while for $J > J_c$
each eigenstate $|\phi_m>$ becomes a mixture of exponentially
many $|\psi_i>$.  It is
convenient to characterize the complexity of an eigenstate $|\phi_m>$
by the quantum eigenstate entropy 
$S_q = -\sum_{i} W_{im} \log_2 W_{im}$, where $W_{im}$ 
is the quantum probability
to find the state $|\psi_i>$ in the eigenstate 
$|\phi_m>$ ($W_{im}=|<\psi_i|\phi_m>|^2$). In this way
$S_q=0$ if $|\phi_m>$ is represented by one $|\psi_i>$,
$S_q=1$ if two $|\psi_i>$ with equal probability
contribute to one $|\phi_m>$ and the maximal
value $S_q=n$ corresponds to a mixture of all $N_H$
states in one $|\phi_m>$. A mixture of two
states $|\psi_i>$ is already sufficient to
modify strongly the quantum register computations
and it is natural to determine the critical coupling
$J_{cs}$ by the condition $S_q(J_{cs})=1$.
The results for $J_{cs}$ dependence on $n$ and $\delta$ are
shown in Figs. 4,5. They clearly show that 
$J_{cs} \approx 0.13 J_c \approx 0.4 \delta/n \gg \Delta_n$.

In the quantum chaos regime at $J > J_c$ one eigenstate
is composed of $\xi$ states $|\psi_i>$ mixed in the
Breit-Wigner width $\Gamma \sim J^2/\Delta_c \sim J^2n/\delta$.
As for the TBRIM the IPR is exponentially large
and is given by (\ref{xi}) \cite{note}.
For $J > \delta$ the interaction becomes too strong
and $\Gamma \sim J \sqrt{n}$ \cite{gs2}. The pictorial
image of the quantum computer melting is shown in Fig. 6.
For $J > J_c$ the eigenstates become very complex
and the quantum computer hardware and its operability
are destroyed by residual inter-qubit interaction.
 
\begin{figure}
\epsfxsize=3.0in
\epsfysize=3.0in
\epsffile{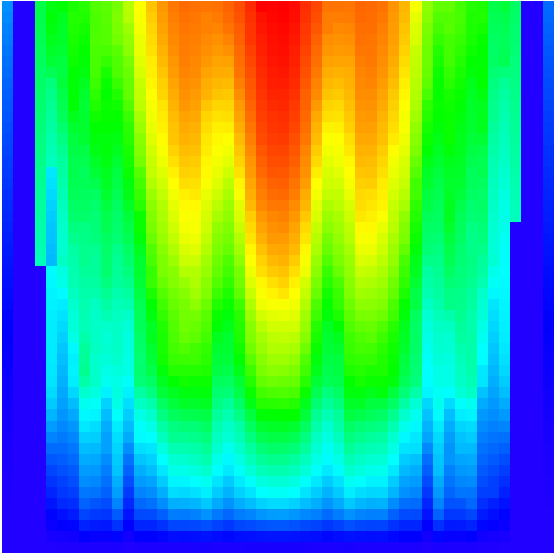}
\vglue 0.2cm
\caption{The quantum computer melting induced by the coupling between qubits.
Color represents the level of quantum eigenstate entropy $S_q$, from bright red
($S_q \approx 11$) to blue
($S_q =0$).  Horizontal axis is the energy of the computer eigenstates 
counted
from the ground state to the maximal energy ($\approx 2n\Delta_0$).
Vertical axis is the value of $J/\Delta_0$, varying from $0$ to $0.5$. Here
$n=12$, $\delta=\Delta_0$, $J_c/\Delta_0=0.273$, and
one random realization of (\ref{hamil}) is chosen.
After [24].
} 
\label{fig6}
\end{figure}

\section{Time scales for quantum chaos and decoherence
in quantum computing}

The results of above section definitely show
that the optimal regime for quantum computing 
corresponds to the integrable region below chaos 
border $J < J_c$. However, it is possible that
for certain experimental proposals it will be
difficult to avoid the quantum chaos regime $J > J_c$.
Therefore, it is also important
to understand during what time scale the quantum chaos
becomes completely developed. As for TBRIM this chaotic
time scale $\tau_{\chi}$ is given by the decay rate
$\Gamma$ from one quantum register state to all others
\cite{gs1,gs2}:
\begin{equation}
\label{tau}
\tau_{\chi} \approx 1/ \Gamma, \;\; \Gamma \sim J^2 n/\delta, 
\;\; \Gamma \sim J \sqrt{n}
\end{equation} 
where the expressions for $\Gamma$ are given for $J < \delta$
and $J > \delta$ respectively. For $\delta=\Delta_0$ a similar
estimate for $\tau_{\chi}$ was given in \cite{flambaum2}. 

The detailed numerical studies of chaotic disintegration of 
an initial state $|\chi (t=0)>=|\psi_{i_0}>$, corresponding 
to the quantum register state $i_0$, are done in \cite{gs2}.
There for $J > J_c$ the behavior  of  the projection probability 
$F_{ii_0} (t) = |<\psi_i|\chi (t)>|^2$ is found and it is
shown that the probability to stay at the initial
state $F_{i_0i_0}(t)$ decays rapidly to zero with the time scale 
$\tau_{\chi} \approx 1/\Gamma$ (see Fig. 14 there).
The growth of the quantum entropy 
$S(t) = -\sum_{i} F_{ii_0}(t) \log_2 F_{ii_0}(t)$
with time is shown in Fig. 7. It clearly shows that 
after a finite time comparable with $\tau_{\chi}$
exponentially many states are excited and the computer
operability is quickly destroyed. 
\begin{figure}
\epsfxsize=3.2in
\epsfysize=2.6in
\epsffile{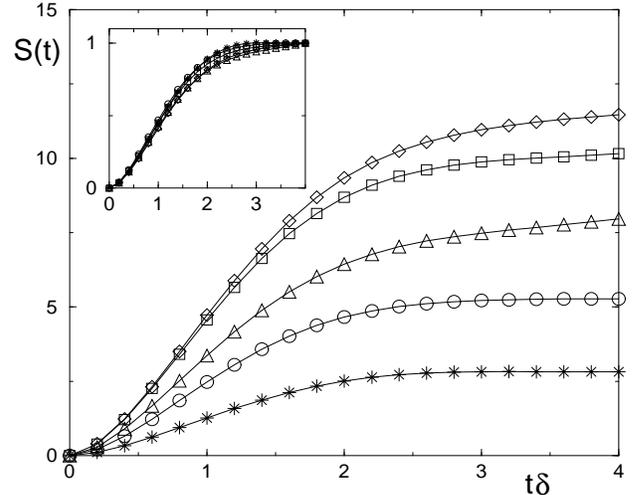}
\caption{Time-dependence of the quantum entropy $S(t)$ for
 $J/\delta=0.4 > J_{c}/\delta$ and $n=16$ (diamonds),
$n=15$ (squares), $n=12$ (triangles), $n=9$ (circles),
$n=6$ (*). Average is made over 200 initial states $i_0$
randomly chosen in the central band. 
Insert shows the same curves normalized to
their maximal value.
After [25].
} 
\label{fig7}
\end{figure}

Above we discussed the emergence of quantum chaos
induced by inter-qubit coupling in an isolated quantum computer.
It is possible to assume that this process can model
to a certain extend the effects of decoherence 
induced by coupling to external world. Indeed, 
the two-body nature of interaction is also valid for 
external coupling. In addition it is also important to
discuss another type of danger for quantum computing
which is not necessary related to mixing and complex
structure of eigenstates. Indeed, the algorithms
constructed for quantum computing, e.g. Shor's or
Grover's algorithms, are based on ideal qubits 
which have $\delta=0$. However, for $\delta > 0$,
and even if $J=0$, there is a phase difference $\Delta \phi$
between different states in one band which
grows with  time as $\Delta \phi \sim \Delta E \; t$,
where $\Delta E$ is the energy difference between 
states in the band and its maximal value can be estimated as
$\Delta E \sim \delta \sqrt{n}$.
It is natural to assume that as soon as the accumulated
phase difference becomes comparable to 1 the 
computational errors become too large for
correct quantum simulations. Therefore, in addition
to the chaotic time scale $\tau_{\chi}$, there exists
the phase coherence time scale $\tau_{\phi}$
which is determined by the condition 
$\Delta \phi(\tau_{\phi}) \sim \Delta E \tau_{\phi} \sim 1$.
This scale is finite even at $J=0$ and can be estimated
as $\tau_{\phi} \sim 1/(\delta \sqrt{n})$ since 
the band width is $\Delta E \sim \delta \sqrt{n}$
and the computer usually operates with all states inside
the band. For small $J$ this estimate is still valid
and in this case $\tau_{\phi} < \tau_{\chi}$
for $J < \delta/n^{1/4}$.
On the contrary for $\delta =0$ and $J > 0$
the energy band width is $\Delta E \sim J \sqrt{n}$
and according to (\ref{tau}) both scales are
comparable $\tau_{\phi} \sim \tau_{\chi}$.
While both time scales $\tau_{\chi}$ and $\tau_{\phi}$
are important for the quantum computer operability
it is clear that the effects of quantum chaos
on the scale $\tau_{\chi}$ are much more dangerous
since after this scale an exponentially large
number of quantum register states are mixed
as it is shown in Fig. 7.
On the contrary on the scale $\tau_{\phi}$
only the phases are changed but not the number of states.
Therefore, it is natural to expect that phase spreading
can be more easily suppressed by error-correcting
codes than the onset of quantum chaos.
\begin{figure}
\epsfxsize=3.2in
\epsfysize=2.6in
\epsffile{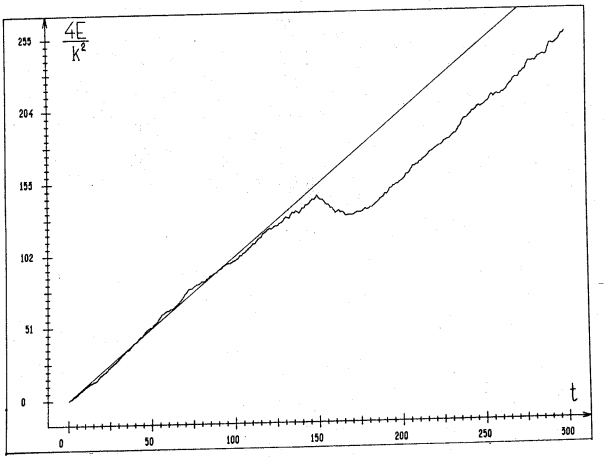}
\caption{Time-dependence of the energy 
$E=<n^2/2>$ in the Chirikov standard
map (\ref{stmap}) in the chaotic regime with
$K=5, \; k=20$ for 1000 orbits
homogeneously distributed at $n=0$ for $t=0$. 
The straight line shows the theoretical
diffusion $E=k^2t/4$. All velocities are inverted
after t=150 with computer accuracy $10^{-12}$
that destroys time reversibility after 20 iterations.
After [53].
} 
\label{fig8}
\end{figure}

These results show that there are two important
time scales $\tau_{\chi}$ and $\tau_{\phi}$.
The effect of external noise and decoherence can
be also characterized by two different scales.
Indeed, from one side the noise gives some effective
rate $\Gamma_T$ with which a multi-qubit state
decays to other states that determines the time scale
$\tau_{\chi} \sim 1/\Gamma_T$. 
It is clear that $\Gamma_T \propto n$
since the noise acts on all qubits
independently. At the same time
the noise gives some effective diffusion
in energy which can be estimated as
$D_E = \omega^2 \Gamma_T$ where $\omega $
is a typical energy change induced by noise
during time $1/\Gamma_T$. For example, noisy
fluctuations of $J_{ij}$ in (\ref{hamil})
give transitions with an energy change
$\omega \sim \delta$. As a result,
$\Delta E \sim \sqrt{D_E \tau_{\phi}}$
and $\Delta \phi \sim {D_E}^{1/2} {\tau_{\phi}}^{3/2} \sim 1$.
This determines two time scales for external decoherence:
\begin{equation}
\label{noise}
\tau_{\chi} \approx 1/ \Gamma_T, \;\; 
\tau_{\phi} \approx 1/ (\omega^2 \Gamma_T)^{1/3}
\end{equation} 
where $\tau_{\chi}$ is related to the decay and relaxation,
while $\tau_{\phi}$ determines the phase coherence.
Let us note that the phase decoherence was discussed for
electrons in metals at low temperature \cite{aronov}
and was observed experimentally, see e.g.  \cite{devoret}. However,
there $\Gamma_T$ was independent of number of electrons,
while for quantum computing $\Gamma_T$ is proportional
to the number of qubits since the global coherence
of multi-qubit states should be preserved.
\begin{figure}
\epsfxsize=3.2in
\epsfysize=2.6in
\epsffile{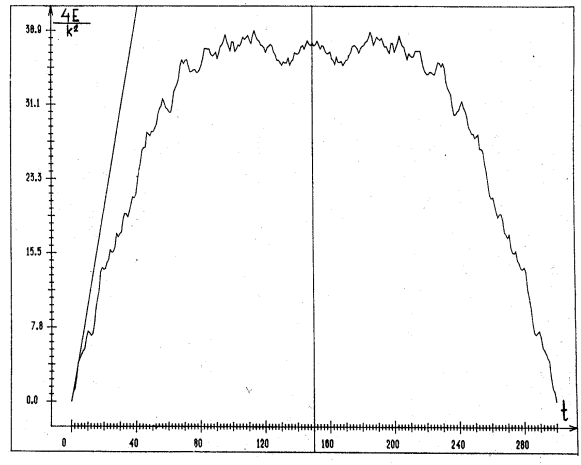}
\caption{Same as in Fig. 8 but for the corresponding
quantum system (\ref{kr}) with
$K=5, \; k=20, \hbar=1$ realized on the same computer.
The vertical line marks the inversion of time
made by replacement $\psi \rightarrow \psi^*$.
The quantum dynamics remain perfectly reversible.
After [53].
} 
\label{fig9}
\end{figure}

The time scales (\ref{noise}) play an important role for
quantum computing. Indeed, the gate operations
should be fast enough comparing to these scales to allow
to realize error-correcting codes \cite{shor2,steane1}
and to avoid a destruction of operability after 
the time scales (\ref{noise}). Here, it is important to
stress an important property of quantum evolution 
for which there is no exponential growth of errors,
contrary to the classical dynamics. An illustration
of this fact is based on the dynamics of the Chirikov
standard map \cite{chirikov79}:
\begin{equation}
\label{stmap}
{\bar n} = n +k\sin(\theta +Tn/2), \;\; 
{\bar \theta} = \theta + T(n + {\bar n})/2
\end{equation} 
Here, $(n, \theta)$ is a pair of momentum and phase variables
and bar denotes the new values of variables after
one period of perturbation. The classical dynamics
depends only on the chaos parameter $K = kT$
and for $K>0.9716..$ the dynamics becomes globally diffusive
in $n$. In this regime the classical trajectories are
exponentially unstable and have positive Kolomogorov entropy.
Due to that the computer errors grow exponentially 
quickly in time that in practice 
destroys the time-reversibility of the map (\ref{stmap}).
This fact is illustrated on Fig. 8 \cite{dls83} where the iterations 
are done on a computer with errors of the order of $10^{-12}$.
Due to exponential instability of classical orbits
the time-reversibility for orbits with inverted momentum
($n \rightarrow -n$) completely disappears after 
20 iterations. The situation is absolutely different
for the corresponding quantum dynamics
described by the unitary evolution operator for wave
function $\psi$ on one iteration \cite{dls83}:
\begin{equation}
\label{kr}
{\bar \psi} = \exp(-iT{\hat n}^2/4) \exp(-ikcos \theta)  \exp(-iT{\hat n}^2/4)
\end{equation} 
where ${\hat n} = -i d/ d \theta$ and $\hbar=1$.
Here the quantum dynamics simulated on the same computer 
remains exactly reversible as it is shown in Fig. 9.

The above results clearly show that there is no 
exponential instability in quantum mechanics \cite{dls83}.
This gives a direct indication that the errors in quantum computing
can be efficiently corrected since e.g. Shor's algorithm
requires relatively small number of gate operations $N_G$
for number factorization: $N_G \sim 300 L^3$ where
$L$ is a number of digits in the number to be factorized \cite{shor1}.
In a sense the Shor algorithm is exponentially fast
while the errors grow only as a power of time that opens 
broad perspectives for quantum computing. This however requires
a development of efficient error-correcting codes.
There is also another general question related to
the uncertainty relation between energy and time. Indeed,
the quantum computing during time $\Delta t$
allows to resolve energy levels only on the energy scale
$\Delta E \sim 1/\Delta t$. This means that the resolution
of exponentially small spacing $\Delta E \sim \Delta_n$
requires exponentially long time analogous to the Heisenberg
time scale in the quantum chaos $t_H \sim 1/\Delta_n$.
Apparently only after this time scale all exponentially
large information hidden in the Hilbert space can become
available. However, it is possible that on shorter time
scales a useful information can be extracted in a way
unaccessible to classical computers, e.g. Shor's
factorization. But a question about how many such
exponentially efficient algorithms can be found
remains open.

\section{Conclusion}

In this paper the conditions for emergence of quantum chaos,
ergodicity and dynamical thermalization in many-body 
quantum systems with interaction and disorder 
are presented. The {\AA}berg criterion represents the 
main condition for onset of quantum chaos and different checks 
performed for various physical systems confirm its
validity. The  generalization of these results
allows to determine the quantum chaos border for quantum
computing. In particular, it is shown that
the critical coupling between qubits,
which leads to quantum chaos and quantum computer hardware melting, 
drops only linearly
with the number of qubits and is exponentially 
larger than the energy level spacing between 
eigenstates of quantum computer.
In this sense the ideal multi-qubit structure
is rather robust in respect to perturbations 
that opens broad possibilities for realization
of quantum computers.

Of course, the optimal regime for quantum computer operability
corresponds to the integrable regime below the quantum chaos
border. In this respect the quantum chaos is a negative 
effect for quantum computing which should be eliminated.
Here it is possible to make an analogy with the
development of classical chaos. This phenomenon also
has negative effects for operability of plasma traps
and accelerators. In fact the very first 
Chirikov resonance-overlap criterion
had been invented in the pioneering work \cite{chirikov59}
for the explanation of experiments on plasma confinement in
open magnetic traps and later found broad applications
in accelerator physics \cite{chirikov69}. Since 1959
it is the only simple physical criterion which allows to determine 
the chaos border in classical nonlinear hamiltonian systems
\cite{chirikov59,chirikov69,chirikov79}
and it merits to be marked by the Nobel symposium organizers.
Indeed, the interest to the  quantum chaos appeared only
after a deep understanding of classical chaos had been reached.
In a similar way it is possible to think that the deep
understanding of quantum chaos in many-body systems
in future will allow to  quantum computers operate in 
a better way.    

The majority of results on quantum chaos in quantum computing
presented in this paper
were obtained with Bertrand Georgeot and it is my pleasure
to thank him for the fruitful collaboration. I am also
thankful to Oleg Sushkov for valuable discussions of 
various questions of interaction and disorder.
I thank the IDRIS in Orsay and the CICT in Toulouse 
for access to their supercomputers. 
This research is partially done in the frame of EC program RTN1-1999-00400.


\vskip -0.5cm
\begin{thebibliography}{99}
\bibitem[*]{byline1} http://w3-phystheo.ups-tlse.fr/$\sim$dima
\bibitem{feynman} R.~P.~Feynman,
   Found. Phys. {\bf 16}, 507 (1986).
\bibitem{steane2}  A. Steane, Rep. Progr. Phys. {\bf 61},
117 (1998).
\bibitem{deutsch} D.~Deutsch, Proc. R. Soc. London Ser. A {\bf 425}, 73 (1989).
\bibitem{divi2} D.~P.~Di~Vincenzo, Phys. Rev. A {\bf 51}, 1015 (1995).
\bibitem{shor1} P.~W.~Shor, in Proc. 35th Annu. Symp. Foundations of
Computer Science (ed. Goldwasser, S. ), 124 (IEEE Computer Society, Los
Alamitos, CA, 1994).
\bibitem{grover} L.~K.~Grover, Phys. Rev. Lett. {\bf 79}, 325 (1997).
\bibitem{shor2} A.~R.~Calderbank  and P.~W.~Shor, 
Phys. Rev. A {\bf 54}, 1098 (1996).
\bibitem{steane1} A.~Steane,
Proc. Roy. Soc. Lond. A {\bf 452}, 2551 (1996).
\bibitem{zoller} J.~I.~Cirac and P.~Zoller,
Phys. Rev. Lett. {\bf 74}, 4091 (1995).
\bibitem{nmr} N.~A.~Gershenfeld and I.~L.~Chuang, 
 Science {\bf 275}, 350 (1997); D.~G.~Cory, A.~F.~Fahmy and T.~F.~Havel,
In Proc. of the 4th Workshop on Physics and Computation
(Complex Systems Institute, Boston, MA, 1996).
\bibitem{vagner} V.~Privman, I.~D.~Vagner and G.~Kventsel, 
Phys. Lett. A {\bf 239}, 141 (1998).
\bibitem{kane} B.~E.~Kane, Nature {\bf 393}, 133 (1998).
\bibitem{bowden} C.~D.~Bowden and S.~D.~Pethel, Int. J. of Laser Phys., to
appear (2000), (quant-ph/9912003).
\bibitem{loss} D.~Loss  and D.~P.~Di~Vincenzo,
 Phys. Rev. A {\bf 57}, 120 (1998).
\bibitem{cooper} Y.~Nakamura, Yu.~A.~Pashkin, and J.~S.~Tsai,  
     Nature {\bf 398}, 786 (1999).
\bibitem{lattice} G.~K.~Brennen, C.~M.~Caves, P.~S.~Jessen and I.~H.~Deutsch
     Phys. Rev. Lett. {\bf 82}, 1060 (1999); D.~Jaksch, H.~J.~Briegel, 
     J.~I.~Cirac, C.~W.~Gardiner and P.~Zoller,
     Phys. Rev. Lett. {\bf 82}, 1975 (1999).
\bibitem{helium} P.~M.~Platzman and M.~I.~Dykman, Science {\bf 284}, 1967
    (1999).
\bibitem{monroe} C.~Monroe, D.~M.~Meekhof, B.~E.~King, W.~M.~Itano
    and  D.~J.~Wineland, Phys. Rev. Lett. {\bf 75}, 4714 (1995).
\bibitem{3q} L.~M.~K.~Vandersypen, M.~Steffen, M.~H.~Sherwood,
    C.~S.~Yannoni, G.~Breyta and I.~L.~Chuang, quant-ph 9910075.
\bibitem{paz} C.~Miquel, J.~P.~Paz and R.~Perazzo, Phys. Rev. A {\bf 54},
    2605 (1996).
\bibitem{zurek} C.~Miquel, J.~P.~Paz and W.~H.~Zurek, 
    Phys. Rev. Lett. {\bf 78},
    3971 (1997).
\bibitem{divi1} D.~P.~Di~Vincenzo, Science {\bf 270}, 255 (1995).
\bibitem{vagner2} D.~Mozyrsky, V.~Privman and I.~D.~Vagner,
    cond-mat/0002350.
\bibitem{gs1} B.~Georgeot and D.~L.~Shepelyansky, quant-ph/9909074,
    to appear in Phys. Rev E.
\bibitem{gs2} B.~Georgeot and D.~L.~Shepelyansky, quant-ph/0005015.
\bibitem{wigner} E.~P.~Wigner, Ann. Math. {\bf 53}, 36 (1951); 
    {\bf 62}, 548 (1955); {\bf 65}, 203 (1957).
\bibitem{mehta} M.~L.~Mehta, Random Matrices (Academic Press,
    Boston, 1991).
\bibitem{guhr} T.~Guhr, A.~M\"uller-Groeling and H.~A.~Weidenm\"uller, 
    Phys. Rep. {\bf 299}, 189 (1999).
\bibitem{tbrim1} J.~B.~French and S.~S.~M.~Wong, Phys. Lett. {\bf 33B},
    447 (1970); {\bf 35B}, 5 (1971).
\bibitem{tbrim2} O.~Bohigas and J.~Flores, Phys. Lett. {\bf 34B},
    261 (1971); {\bf 35B}, 383 (1971).
\bibitem{flambaum1} V.~V.~Flambaum, G.~F.~Gribakin and 
                 F.~M.~Izrailev, Phys. Rev. E
                 {\bf 53}, 5729 (1996); V.~V.~Flambaum, F.~M.~Izrailev 
                 and G.~Casati, {\it ibid.} {\bf 54}, 2136 (1996); 
                 V.~V.~Flambaum and F.~M.~Izrailev, {\it ibid.} 
                 {\bf 55}, R13 (1997).
\bibitem{zelevin} V.~Zelevinsky, B.~A.~Brown, N.~Frazier and M.~Horoi, 
                Phys. Rep.  {\bf 276}, 85 (1996).
\bibitem{berkovits} R.~Berkovits and Y.~Avishai, 
           J. Phys. C {\bf 8}, 391 (1996).
\bibitem{aberg} S.~{\AA}berg, Phys. Rev. Lett. {\bf 64}, 3119 (1990).
\bibitem{aberg1} S.~{\AA}berg, Prog. Part. Nucl. Phys. {\bf 28}, 11 (1992).
\bibitem{dls94} D.~L.~Shepelyansky, Phys. Rev. Lett. {\bf 73}, 2607 (1994).
\bibitem{sushkov} O.P.Sushkov and V.V.Flambaum, 
           Usp. Fiz. Nauk {\bf 136}, 3 
           (1982) [Sov. Phys. Usp. {\bf 25}, 1 (1982)].
\bibitem{dlssush97} D.~L.~Shepelyansky and O.~P.~Sushkov, 
             Europhys. Lett. {\bf 37}, 121 (1997).
\bibitem{weinmann} D.~Weinmann, J.-L.~Pichard and Y.~Imry,
         J. Phys. I France, {\bf 7}, 1559 (1997).
\bibitem{jacquod97} P.~Jacquod and D.~L.~Shepelyansky,
             Phys. Rev. Lett. {\bf 79}, 1837 (1997).
\bibitem{bohigas} O.~Bohigas, M.-J.~Giannoni, and
        C.~Schmit, Phys. Rev. Lett. {\bf 52}, 1 (1984); 
        O.~Bohigas in {\it Les Houches Lecture Series} {\bf 52},
        Eds. M.-J.~Giannoni, A.~Voros, and J.~Zinn-Justin (North-Holland,
        Amsterdam, 1991).
\bibitem{shklovskii} B.~I.~Shklovskii, B.~Shapiro, B.~R.~Sears, 
        P.~Lambrianides and H.~B.~Shore, Phys. Rev. B 
        {\bf 47}, 11487 (1993).  
\bibitem{anote} In fact the border of Anderson transition 
         in 3-dimensions \cite{shklovskii} can be also
         found from the relation (\ref{uc}). Then the average
         square matrix element $U^2/3$ is replaced by
         hopping square $V^2$ while $\Delta_c = W/Z$ where
         $W$ is the amplitude of disorder and $Z=6$
         is the number of coupled sites. Since at the band center 
         the transition happens at 
         $W_c \approx 16.5V$ \cite{shklovskii} we get
         $C \approx 0.63$ being very close to the value in TBRIM.
\bibitem{georgeot} B.~Georgeot and D.~L.~Shepelyansky,
        Phys. Rev. Lett. {\bf 79}, 4365 (1997).
\bibitem{song} P.H.Song, cond-mat/0004237.
\bibitem{mirlin97} A.~D.~Mirlin and Y.~V.~Fyodorov,
        Phys. Rev. B, {\bf 56}, 13393 (1997).
\bibitem{bohr} A.~Bohr and B.~R.~Mottelson, {\it Nuclear Structure}, 
                  Benjamin, New York {\bf 1}, 284 (1969).
\bibitem{glass} B.~Georgeot and D.~L.~Shepelyansky,
                Phys. Rev. Lett. {\bf 81}, 5129 (1998).
\bibitem{note} However, here it is important to note that
        the exponentially large value of IPR $\xi$
        does not imply automatically that the eigenstates
        are chaotic. Indeed, following Shor \cite{shor1},
        one can imagine a perturbation which rotates each qubit
        from a state $|00...0>$ on a random angle.
        In this way exponentially many quantum 
        register states $|\psi_i>$ are present in one new eigenstate
        and exponentially many of them 
        are exited after a finite time of rotation. 
        Nevertheless, these states are not complex/chaotic
        and the level statistics $P(s)$ remains poissonian
        with $\eta=1$.
\bibitem{flambaum2} V.~V.~Flambaum, quant-ph/9911061.
\bibitem{aronov} B.~L.~Altshuler and A.~G.~Aronov, in {\it Electron-electron
        interactions in disordered systems}, Eds. A. L. Efros and M. Pollak,
        North-Holland, Amsterdam (1985), 1.
\bibitem{devoret} A.~H.~Steinbach, J.~M.~Martinis
        and  M.~D.~Devoret, Phys. Rev. Lett. {\bf 76}, 3806 (1996).
\bibitem{chirikov79} B.~V.~Chirikov, Phys. Rep. {\bf 52}, 263 (1979).
\bibitem{dls83} D.~L.~Shepelyansky, Physica D {\bf 8}, 208 (1983).
\bibitem{chirikov59} B.~V.~Chirikov, At. Energ. {\bf 6}, 630 (1959) 
     [Engl. Transl., J. Nucl. Energy Part C: Plasma Phys.
     {\bf 1}, 253 (1960)]. 
\bibitem{chirikov69} B.~V.~Chirikov, 
     Research concerning the theory of nonlinear resonance and stochasticity,
     Preprint N 267, Institute of Nuclear Physics, Novosibirsk (1969) 
     [Engl. Trans., CERN Trans. 71-40 (1971)]

\end{thebibliography}
\end{document}